\documentclass[12pt]{book}

\usepackage[dvips]{graphicx,color}
\usepackage{makeidx,tsukuba}
\makeauthorindex
\makeindex

\begin{document}

\BookTitle{\itshape The 28th International Cosmic Ray Conference}
\CopyRight{\copyright 2003 by Universal Academy Press, Inc.}
\pagenumbering{arabic}

\chapter{TeV Observations of the Galactic Center }

\author{%
%
%
K.~Kosack $^1$,~~$^2$ I.H.~Bond, P.J.~Boyle, S.M.~Bradbury, J.H.~Buckley,
D.~Carter-Lewis, O.~Celik, W.~Cui, M.~Daniel, M.~D'Vali,
I.de~la~Calle~Perez, C.~Duke, A.~Falcone, D.J.~Fegan, S.J.~Fegan,
J.P.~Finley, L.F.~Fortson, J.~Gaidos, S.~Gammell, K.~Gibbs,
G.H.~Gillanders, J.~Grube, J.~Hall, T.A.~Hall, D.~Hanna, A.M.~Hillas,
J.~Holder, D.~Horan, A.~Jarvis, M.~Jordan, G.E.~Kenny, M.~Kertzman,
D.~Kieda, J.~Kildea, J.~Knapp, H.~Krawczynski, F.~Krennrich,
M.J.~Lang, S.~LeBohec, E.~Linton, J.~Lloyd-Evans, A.~Milovanovic,
P.~Moriarty, D.~Muller, T.~Nagai, S.~Nolan, R.A.~Ong, R.~Pallassini,
D.~Petry, B.~Power-Mooney, J.~Quinn, M.~Quinn, K.~Ragan, P.~Rebillot,
P.T.~Reynolds, H.J.~Rose, M.~Schroedter, G.~Sembroski, S.P.~Swordy,
A.~Syson, V.V.~Vassiliev, S.P.~Wakely, G.~Walker, T.C.~Weekes,
J.~Zweerink \\
{\it \small
(1) Washington University Department of Physics, St.
Louis, MO 63130, USA \\
(2) The VERITAS Collaboration--see S.P.Wakely's paper} ``The VERITAS
Prototype'' {\it from these proceedings for affiliations}
}

\section*{Abstract}
We present the results of 16 hours of ongoing observations of the
galactic center region (including Sagittarius A*) with the Whipple
High Energy Gamma-Ray Telescope.  We apply a data analysis method
optimized for large zenith angle observations on an independent Crab
Nebula data set.  We discuss possible systematic problems associated
with observations of extended sources in the galactic plane.

\section{Introduction}

The Galactic Center is a complicated region containing sources which
show emission in radio through x-ray wavelengths. The region also contains
one of the brightest EGRET unidentified sources (2EG~J1746-2852),
which may to be slightly offset from Sgr A* \cite{dingus}. Recent
X-ray observations (with Chandra) indicate flaring activity,
motivating the possibility that the $3\times 10^6 M_\odot$ black hole
at the position of Sgr A* powers a nearby low-luminosity AGN-like
source. Multi-wavelength observations of the galactic center at TeV
energies might provide data allowing us to test models for off-axis
emission from low-luminosity, low accretion rate sources without the
effects of intergalactic IR absorption.  The galactic center is also
the most likely place to look for gamma rays from neutralino
dark-matter annihilation if the galactic halo has a cusp that extends
down to scales of $<100$ parsecs from the center \cite{buckley}.

We have continued observations of the galactic-center region around
Sgr A* with the Whipple 10 m gamma-ray telescope resulting
in an additional on-source exposure of 16 hours over the 2000 to 2003
observing seasons \cite{kosackICRC}. In independent observations from
1995 to 1997 using a lower resolution camera, the source was not
detected, although a $2.5\sigma$ excess was
observed near the approximate position of Sgr A* after 4.5 hours of
observation \cite{buckleyICRC}.  Though no signal was seen in our 2001
results, we have since re-analyzed all of the data taken with the high
resolution camera.

\section{Method}

Whipple gamma ray data are traditionally taken as a series of 28
minute exposures, each of which is followed by an off-source run which
is offset 30 minutes in right ascension for background subtraction. In
the case of Sgr A*, data was taken off-source before the on-source
observations due to a bright star field in the region $30'$ past the
galactic center's position.  Both on and off source data are analyzed
in the same manner, and Gaussian padding is used to bring the
background noise up to the same level in both runs to take care of
brightness differences. For each event in the data, the \v{C}erenkov
light image of an air shower is characterized by a standard set of
parameters originally defined by A. M. Hillas \cite{hillas}. Cuts on
these parameters allow the rejection of background (e.g. cosmic-ray
induced showers).

Since Sgr A* transits at a zenith angle of about $60^\circ$ at the
latitude of the Whipple Observatory, an analysis technique which
scales with zenith angle and the logarithm of the shower size
(\emph{Zcuts}) was used to correct for the geometry at low elevation.
This technique was outlined previously in the ICRC 2001 proceedings
\cite{kosackICRC}. Cuts were optimized on a large database containing
$\approx 20$ hours of Crab Nebula observations.

\begin{figure}
  \includegraphics[width=7cm]{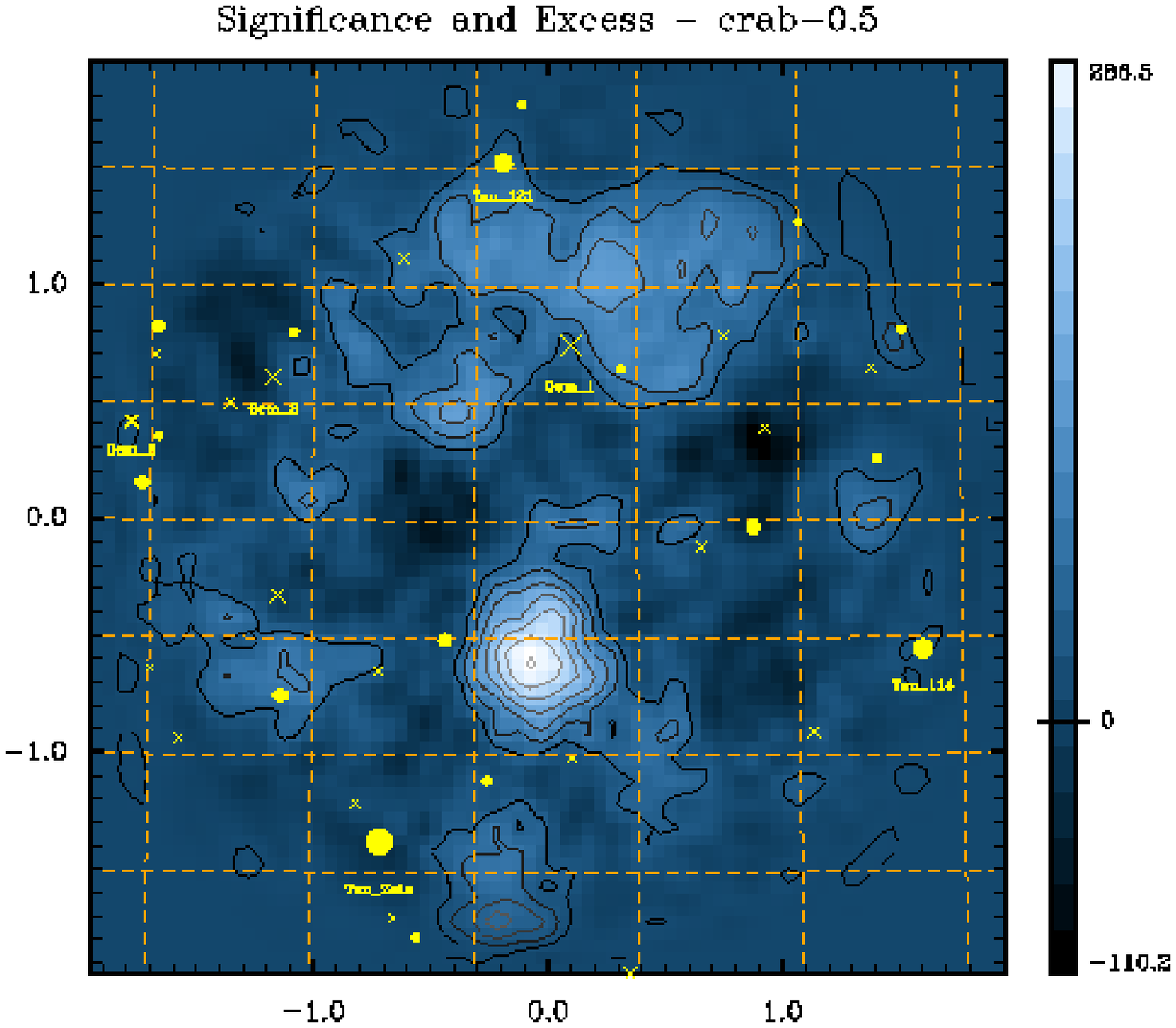}
  \includegraphics[width=7cm]{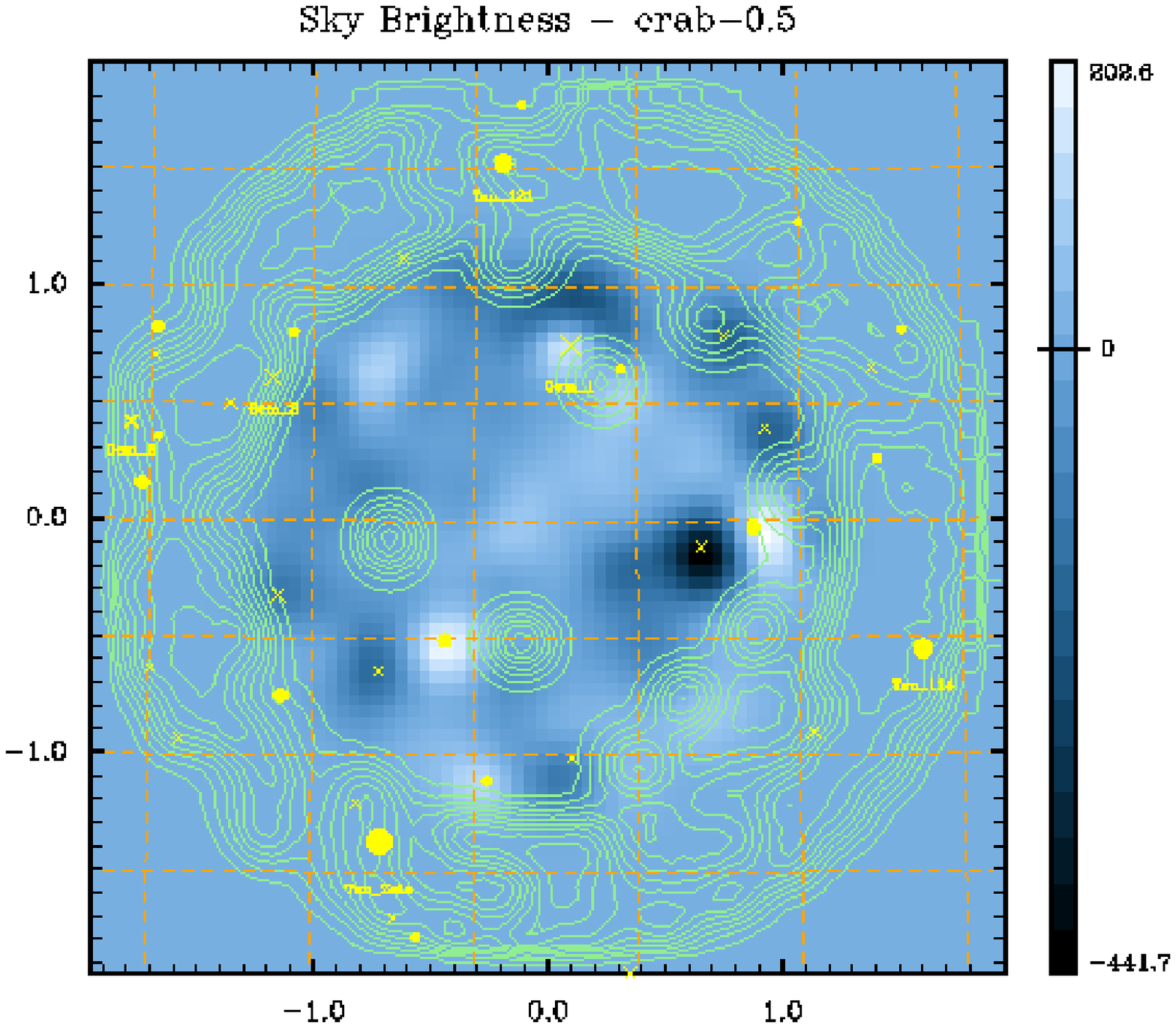}
  \caption{\label{fig:crab} A 2-D gamma ray image of the Crab
  Nebula, offset by -0.5 degrees in declination. The on-source
  exposure time is 2.8 hours.  The left image shows the gamma ray
  excess (color image, with the scale on the right in excess counts)
  with overlaid significance contours ($1\sigma$ per contour). The
  right image is the optical sky brightness map (with arbitrary
  scale), where the contours show regions affected by tubes that have
  been turned off.   RA (vertical dashed)
  and Dec (horizontal dashed) lines are shown. The positions of bright
  stars (dots for on-source, x's for off-source) have been overlaid in
  both plots.  The axes are in degrees from the center of the camera.
  }
\end{figure}

To produce a 2-D image, we reconstruct the two possible points of
origin based on the elongation of the elliptical shower image and
select the correct one based on the shower asymmetry. Analysis
providing a two-dimensional map of the gamma ray emission was required
in this case due to the uncertainty in the exact position of the
source and pointing error when the telescope is at low elevation.  To
test our point-of-origin reconstruction technique, we analyzed data
taken from the Crab Nebula during which the telescope was offset by
$0.5^\circ$ in declination. A plot of data (not from the optimization
set) showing the Crab Nebula offset in the field of view is shown in
Figure \ref{fig:crab}

To accurately determine the telescope's pointing and to characterize
any systematic effects from bright stars, we generate a 2-D brightness
map of the sky using the pedestal variances of each photo-tube in the
camera. This in effect produces a low-resolution optical image of area
of the sky which we then compare to a database of known star
positions.  For further characterization of systematic error, we also
generate a cumulative 2-D plot which shows areas of the field of view
which are affected by photo-tubes which have been turned off (due to a
bright star in the field or malfunction).  An example of these maps is
shown in Figure \ref{fig:crab}

\section{Results}

\begin{figure}
  \begin{small}
  \includegraphics[width=7cm]{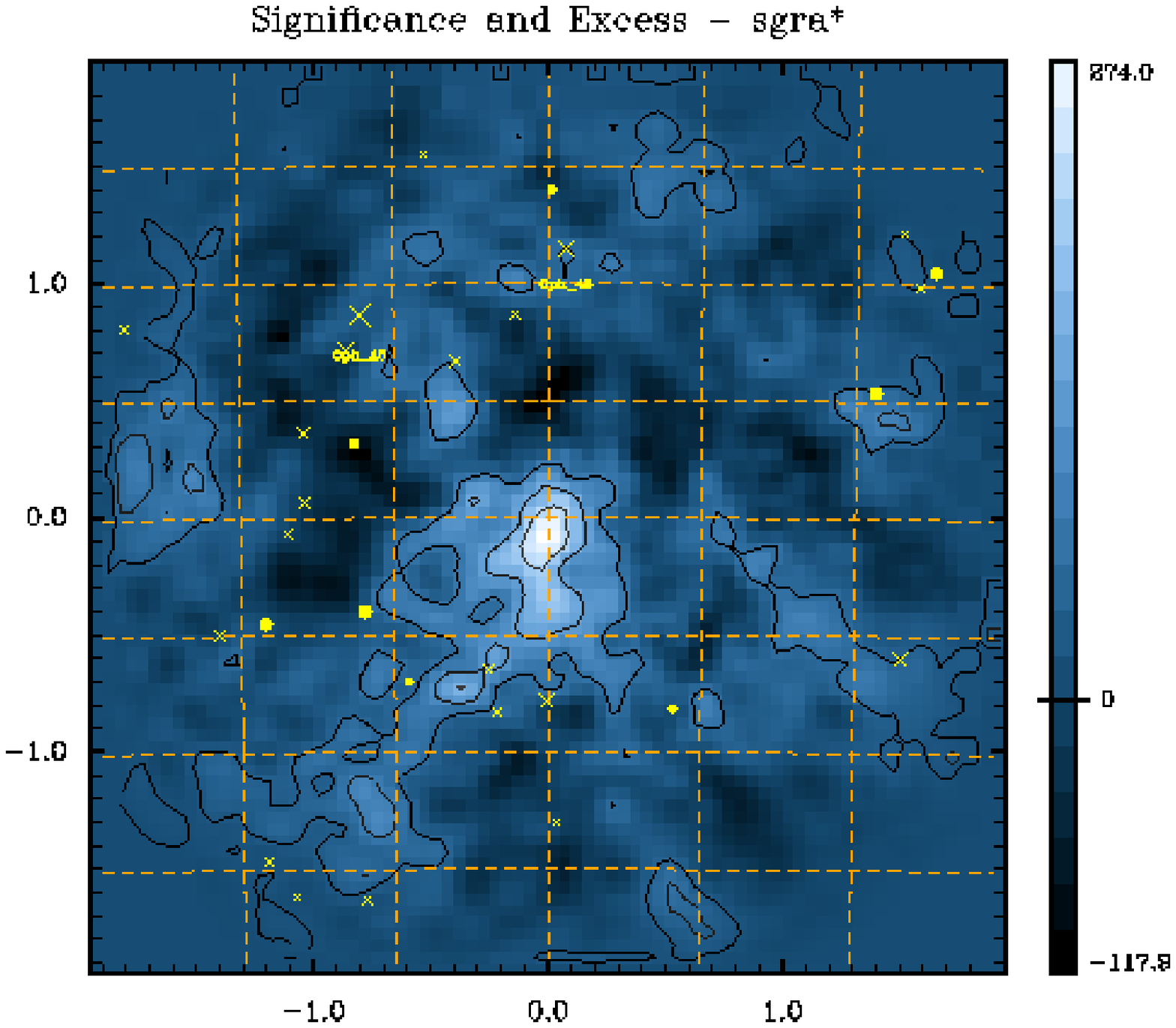}
  \includegraphics[width=7cm]{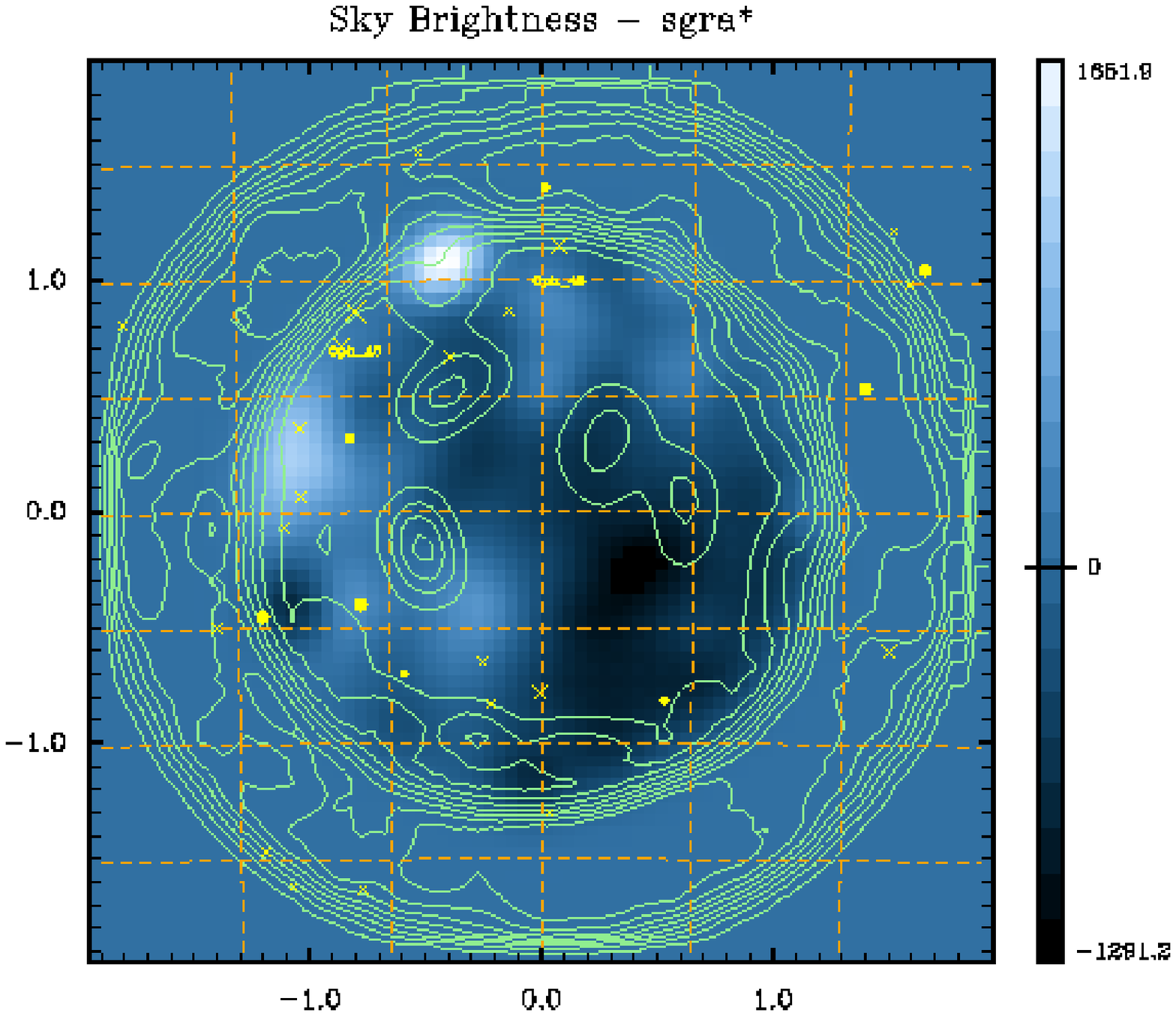}
  \caption{\label{fig:sgra} A 2-D gamma ray image of the region around
  Sgr A*.  The left image shows the gamma ray excess (color image,
  with the scale on the right in excess counts) with overlaid
  significance contours ($1\sigma$ per contour). The right image is
  the optical sky brightness map, where the contours show regions
  affected by tubes that have been turned off.  RA (vertical dashed)
  and Dec (horizontal dashed) lines are shown. The positions of bright
  stars (dots for on-source, x's for off-source) have been overlaid in
  both plots.  The axes are in degrees from the center of the camera.}
  \end{small}

\end{figure}

We have analyzed 16 hours of data taken with the telescope pointed at
Sgr A*.  We see an excess of an excess of $3.7\sigma$ at the center of
the field of view (see Figure \ref{fig:sgra}). The approximate energy
threshold for these observations is $2~TeV$ \cite{petry}.  We
considered three positions in the field of view as potentially
interesting: the position of Sgr A* ($3.7\sigma$), the EGRET source
($1.81\sigma$) \cite{dingus}, and the $2.5 \sigma$ excess seen in
previous observations \cite{buckleyICRC}.  The excess does not take
into account the trials factor associated with these three
possibilities. Significant systematic errors might be present due to
the large zenith-angle of the observations.  Pointing checks show that
there is a $0.1^\circ$ error in the camera's position at low
elevation. To check the significance of the excess, we are in the
process of analyzing the data using other 2-D analysis programs.
Corrections for pointing errors and a detailed spectral analysis of
the data is currently under way and will be presented at the
conference.

\section{Discussion}

More data of the Galactic Center are currently being taken and further
analysis is under way.  If there is a gamma ray source, Southern hemisphere
telescopes with their reduced energy thresholds should be able to
readily detect it, but the LZA observations presented here provide a
larger effective area at high energies. These observations provide
constraints on models for the X-ray flaring, as well as better
constraints on annihilation of high energy neutralinos at the Galactic
Center.

\endofpaper
\end{document}